\documentclass[twocolumn,prl,aps,showpacs]{revtex4}
\usepackage{amsfonts}
\usepackage{amssymb}
\usepackage{epsfig}
\usepackage{amsbsy}
\usepackage{amssymb}
\usepackage{amsmath}
\usepackage{graphicx}
\usepackage{graphics}

\begin{document}

\title{The topological system with a twisting edge band:
position-dependent Hall resistance }

\author{ Xuele Liu$^{1}$}\altaffiliation{Correspondences send to xuele@okstate.edu}
\author{Qing-feng Sun$^{2}$}
\author{X.C. Xie$^{3}$}

\affiliation{ $^1$Department of Physics, Oklahoma State University,
Stillwater, Oklahoma 74078, USA}
\affiliation{$^2$Institute of Physics, Chinese Academy of Sciences, Beijing 100190, China}
\affiliation{$^3$International Center for Quantum Materials, Peking
University, Beijing 100871, China }
\date{\today}

\begin{abstract}
We study a $\nu=1$ topological system with one twisting edge-state
band and one normal edge-state band. For the twisting edge-state
band, Fermi energy goes through the band three times, thus,
having three edge states on one side of the sample;
while the normal edge band contributes only one
edge state on the other side of the sample. In
such a system, we show that it consists of both topologically protected
and unprotected edge states, and as a consequence, its Hall resistance depends on
the location where the Hall measurement is done even for a
translationally invariant system. This unique property is absent in a normal topological insulator.
\end{abstract}

\pacs{73.43.-f, 73.20.-r, 73.23.Ad}
 \maketitle

\emph{Introduction:} The topological system has attracted much
attention in recent years \cite{rv1,rv2}. About twenty years
ago, by proposing the quantum anomalous Hall effect (QAHE) in
graphene \cite{haldane}, Haldane gave a simple two-band model to
study a topological system. Recently, the topological
insulator material is first predicted and then experimentally
observed in some two-dimensional (2D) systems \cite{addref1,2Dth,
2Dex}. The three-dimension topological materials are also discovered soon
after \cite{3D}.

In research of the robustness of topological system, the analysis of
edge states is to be an effective approach \cite{edgestates}. The helical
edge states for 2D topological systems are shown to have the topological
protection of $Z_{2}$ \cite{hedge}, and the scattering between them
is prohibited without breaking time reversal symmetry.
While with edge bands distortion,
they may cross the Fermi surface more than one time, which may also
give rise to some extra edge states \cite{rv1}. However, these extra edge
states can not bring new topological phases, and are not protected by
the topology \cite{fu2007}. They are thought easy to be affected and
are treated as unimportant in the earlier studies.

In this Letter, we show a nontrivial effect from the topological
unprotected edge states. While a system is with both the
topological protected and unprotected edge states,
the Hall conductance depends on the measurement location even
for a translationally invariant system. This novel property
survives at a finite disorder, however, it is absent in both
topological trivial systems and normal topological systems.
Thus, this unique property is the hallmark of a topological system with
a twisting edge band.

\emph{Model and Hamiltonian:} The band structure of our system
is shown in Fig.1(A). Below we provide one example
of how to achieve this band structure. Without loss of generality, we take
the simple $\nu=1$ topological system as an example, which consists of
one pair of topological protected edge states. The AB-stacked square
lattice QAHE system \cite{liu2011} is chosen, in which the two type
of atoms are needed. As shown in Fig. 1(C), we can assume atom $A$
at $s$ level and atom $B$ at the lowest $p$ level \cite{wu2010}.
Generally, this $p$-orbital may not along the direction of lattice
structure, here we choose it along $\pm\vec{e}_{1}$-direction. The
check board magnetic field is also applied by the Peierls phase
$\phi_{0}=\pi/2$ when an electron jumps from $A$ to $B$ along $\pm
\vec{e}_{y}$-direction. Supposing the on-site energy of $A$ and $B$
are the same, set to be the zero energy point. The tight-binding
Hamiltonian can thus be written as $H =H_{1}+H_{2}$, with $H_{1}$
($H_2$) the nearest (next-nearest) hopping Hamiltonian:
\begin{eqnarray}
  \nonumber H_{1} &=& -t_{ab}\sum_{\mathbf{i}}\left[b_{\mathbf{i}+\delta x}^{\dag}a_{\mathbf{i}}+{\mathrm e}^{{\mathrm i}\phi_{0}}b_{\mathbf{i}+\delta y}^{\dag}a_{\mathbf{i}}+h.c.\right]\\
   && +t_{ab}\sum_{\mathbf{i}}\left[a_{\mathbf{i}+\delta x}^{\dag}b_{\mathbf{i}}+{\mathrm e}^{-{\mathrm i}\phi_{0}}a_{\mathbf{i}+\delta y}^{\dag}b_{\mathbf{i}}+h.c.\right]\\
   \nonumber H_{2} &=& -\sum_{\mathbf{i}}\left[t_{a1}a_{\mathbf{i}+\delta e1}^{\dag}a_{\mathbf{i}}+t_{a2}a_{\mathbf{i}+\delta e2}^{\dag}a_{\mathbf{i}}+h.c.\right]\\
    &&-\sum_{\mathbf{i}}\left[t_{b1}b_{\mathbf{i}+\delta e1}^{\dag}b_{\mathbf{i}}+t_{b2}b_{\mathbf{i}+\delta e2}^{\dag}b_{\mathbf{i}}+h.c.\right]
\end{eqnarray}
Here $t_{ab}$ is the hopping term from $A$ to $B$ along the $+x$ and
$+y$-directions, set to be positive; while the hopping from $B$ to
$A$ along the same direction get a negative sign. $t_{a1}$ and
$t_{a2}$ are the next nearest neighbor hopping at $A$ sublattice along
$\vec{e}_{1}$ and $\vec{e}_{2}$, respectively. $t_{b1}$ and $t_{b2}$
are the counterparts for the $B$ sublattice. One can check
$t_{a1},t_{a2},t_{b2}>0$ and $t_{b1}<0$ from Fig. 1(C). Besides, we
also have $t_{a1}=t_{a2}$ and $|t_{b1}|\neq|t_{b2}|$, due to the
anisotropy of the $p$ level.

It is easy to discuss this tight-binding Hamiltonian in
$\mathbf{k}$-space. Because the system is translationally invariant,
we have
$\mathcal{H}(\mathbf{k})=h_{0}(\mathbf{k})+\mathbf{\sigma}\cdot
\mathbf{p}(\mathbf{k})$. Here
$p_{x}(\mathbf{k})=2t_{ab}\sin(k_{y}a_{0})$,
$p_{y}(\mathbf{k})=-2t_{ab}\sin(k_{x}a_{0})$. The next nearest
hopping gives $p_{z}(\mathbf{k})=-
[(t_{a1}-t_{b1})\cos(k_{x}a_{0}+k_{y}a_{0})+(t_{a2}-t_{b2})\cos(k_{x}a_{0}-k_{y}a_{0})]$
and a nonconstant
$h_{0}(\mathbf{k})=-[(t_{a1}+t_{b1})\cos(k_{x}a_{0}+k_{y}a_{0})+(t_{a2}+t_{b2})\cos(k_{x}a_{0}-k_{y}a_{0})]$.
Here $a_{0}$ is the distance between the nearest neighbor atoms $A$ and $B$. The
Chern number of the system can be calculated in $\mathbf{k}$-space
\cite{rv1,TKNN} by $\nu=\int d^{2}\mathbf{k}\mathcal{F}/2\pi$. For
our system, when there exists a real gap, the Chern number of the
lower band gives $\nu=1$.

\begin{figure}
[t]
\begin{center}
\includegraphics[width=8.2cm]{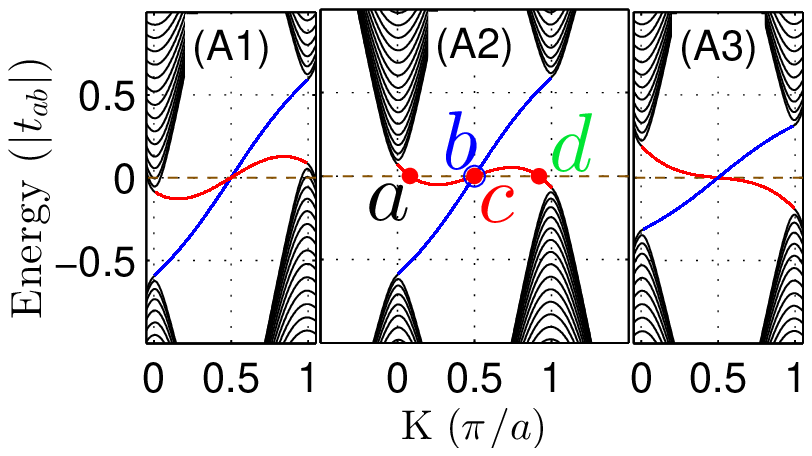}\\
\includegraphics[width=4.2cm]{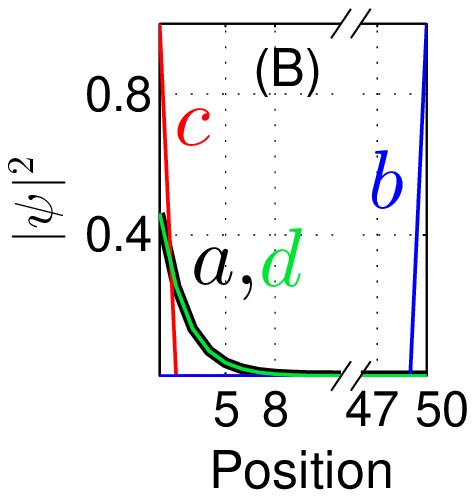}\hspace{0.3cm}\includegraphics[width=4cm]{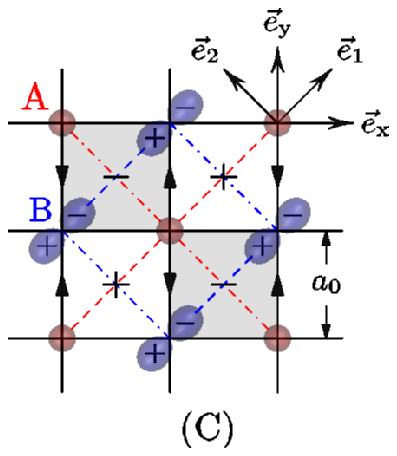}\\
\vspace{0.2cm}
\includegraphics[width=4.35cm]{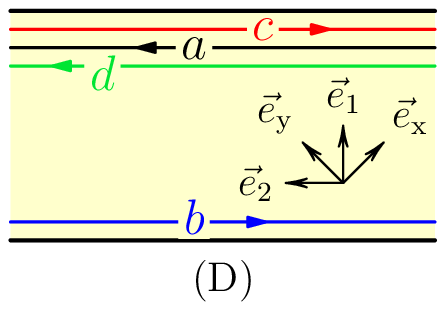}\hspace{0.35cm}\includegraphics[width=3.6cm]{1E.eps}
\end{center}
\caption{\label{fig:Fig. 1} (Color online) (A) The energy band
structures of zigzag-edge ribbon of topological
system, with the ribbon width $W=50a$ 
and $a=\sqrt{2}a_{0}$. We choose the parameters $t_{ab}=10$ and
$t_{a1}=t_{a2}=t_{c}+0.1$, $t_{b1}=-t_{c}-t_{s}$, $t_{b2}=t_{c}$ for
all the subplots. (A1) indirect semi-metal with $t_{c}=1.4$ and
$t_{s}=-0.4$, (A2) the twisting edge band system with $t_{c}=1.4$
and $t_{s}=0.4$, and (A3) the normal topological system with
$t_{c}=0.7$ and $t_{s}=1$. (B) The distribution $|\psi|^2$ of the
four edge states of (A2). (C) The lattice structure of the system.
(D) The schematic diagram of the four edge states of (A2). (E) The
twisting edge band of (A2) can be treated as mix of the topological
protected and unprotected systems.}
\end{figure}

\emph{Twisting edge band:} The coexistence of distorted edge band
and normal edge band originates from the symmetry breaking of the
eigenvalue $\lambda_{\pm}=h_{0}\pm |\mathbf{p}|$. These two
eigenvalue correspond separately to the upper and
down bands. Because of the next nearest hopping, the symmetry
of $\lambda_{\pm}$ reduces from $C_{4}$ to $C_{2}$. The Dirac points
at $(0,0)$ and $(\pm \pi,\pm \pi)$ have different energy values as
the Dirac points at $(0,\pm \pi)$ and $(\pm \pi,0)$.
If the system is constrained at $\vec{e}_{x}$ or $\vec{e}_{y}$
direction, each projected Dirac point in fact contains two type of
Dirac points, so the projected Dirac points remains the same.
However, if the system is constrained at $\vec{e}_{1}$-direction
[see Fig.1(C)], i.e., if with the zigzag edge, each projected
Dirac point contains only one type of Dirac point, the two projected
Dirac points are different with each other, as shown in Fig.1(A).
Consequently, the two edge bands may have different group velocities
$|\partial \varepsilon (\mathbf{k})/\partial \mathbf{k}|$. In this
way, the edge bands are distorted.

To get a twisting edge band, we need a little more effort. Define
$A_{t}=(t_{a1}+t_{a2})$ and $B_{t}=(t_{b1}+t_{b2})$, we can get the
bulk gap of the system as $\Delta=2(|A_{t}-B_{t}|-|A_{t}+B_{t}|)$.
While $A_{t}B_{t}>0$ gives an indirect negative gap $\Delta$. In
this case, although the system has a twisting edge band [see red
curve in Fig.1(A1)], but it is without a bulk gap, which creates
an indirect semi-metal. The bulk insulator needs $\Delta>0$ thus
$A_{t}B_{t}<0$. When gap $\Delta$ is large, the system may only have
a distorted edge band but no twisting edge band [see Fig.1(A3)],
which is the normal 2D topological insulator. When gap $\Delta$ is
positive but small, we may have a twisting edge band [Fig.1(A2)]. We
also have another bigger `gap'
$\Delta_{2}=2(|A_{t}-B_{t}|+|A_{t}+B_{t}|)$, corresponding to the
normal edge band [see the blue curve in Fig.1(A)]. In our system
$A_{t}>0$, and it's no harm to set $A_{t}>|B_{t}|$, then we can get
$\Delta=-4B_{t}$ and $\Delta_{2}=4A_{t}$. This means that the
twisting and normal edge bands are independently determined by
$B_{t}$ and $A_{t}$, respectively. If we choose
$t_{b1}=-t_{c}-t_{s}$, $t_{b2}=t_{c}$ with $t_{c},t_{s}>0$, the gap
$\Delta$ is simplified to $\Delta=4t_{s}$.

Due to the edge band being twisted, it can cross Fermi surface
$E_{F}$ three times, marked by $a$, $c$ and $d$ [see
Fig.1(A2)]. The other normal edge band meets Fermi surface at $b$.
Fig.1(B) shows the distribution $|\psi|^2$ v.s. location for these
four states. We can see that, all four states are localized on
the edges of the sample, the distribution is almost zero inside the
bulk. Among them, the three states $a,c$ and $d$ are localized on
the upper edge, while state $b$ is localized on the lower edge [see
also Fig.1(D)]. As the Chern number of the system is $\nu=1$, only
one pair of the edge states are protected by the topology, while
another two are not protected. There is no doubt that $b$ is protected
by topology since it is the only one edge state on the lower edge.
The other topology protected state is a mixture of these
three degenerated edge states $a,c$ and $d$ on the upper edge. Here
we notice that the present system can be treated as the combination
of the normal topological system plus a topological trivial system
with one pair of unprotected edge states, as shown in
Fig.1(E).

\emph{The translational invariance symmetry breaking of the Hall
resistance.} Now, let us study the transport property of the
system using the $6$-lead set-up. As shown in Fig.2(c), lead-1 and
lead-4 are made by the same materials of the sample, which can
support the well-defined edge states inside the gap of sample. The
vertical leads $2,3,5,6$ are made of a metal, which can afford as much
modes as possible. A small longitudinal voltage gradient is applied
by setting the lead-1 at $V/2$ and the lead-4 at $-V/2$, providing the
longitudinal current $I_{1}$.
We use the zero
temperature Landauer-B$\ddot{u}$ttiker formula
$I_{p}=\frac{e^2}{h}\sum_{q\neq p}(V_{p}-V_{q})T_{p,q}$,
with $T_{p,q}$ the transmission coefficient from the lead $q$ to
$p$ \cite{addref2}. The vertical voltage $V_{p}$ can thus be obtained by
using the open boundary condition, i.e. by letting the corresponding
leads to have zero current: $I_{p}=0$ with $p=2,3,5,6$. Finally, the
Hall and longitudinal resistances can be obtained from $R_{p,q}
\equiv (V_{p}-V_{q})/I_{1}$.

\begin{figure}
[t]
\begin{center}
\includegraphics[width=8.2cm]{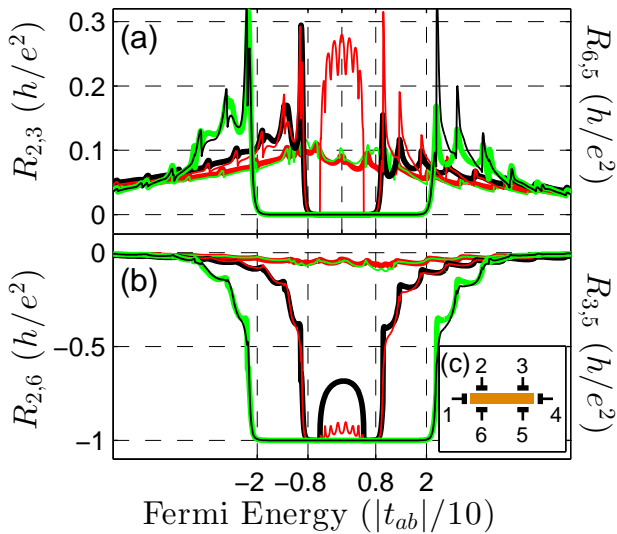}
\end{center}
\caption{\label{fig:Fig. 2} (Color online) For the three sets of
parameters used in Fig.1(A), the corresponding resistances of the
system v.s. the Fermi energy: (a) the longitudinal resistances and
(b) the Hall resistances. The wide lines are for $R_{6,5}$ and
$R_{2,6}$, the narrow lines are for $R_{2,3}$ and $R_{3,5}$. In both
figures, the pair of lines with the broadest quantized plateau
($-2\sim2$) are for Fig.1(A3); the pair of lines only have plateau
within $-0.8\sim0.8$ are for Fig.1(A2); for Fig.1(A1), the pair of
lines have no plateau. Other parameters used for the calculation:
the ribbon width $W=50a$, the distance between vertical leads $L=
20a$. (c) is the schematic diagram of the $6$-lead measurement we
used for (a) and (b).}
\end{figure}

For the three sets of parameters used in Fig.1(A), by changing the
Fermi energy, in Fig.2(b) we plot the Hall resistance $R_{2,6}$,
measured on the left side of the sample, and $R_{3,5}$,
on the right side. We also draw in Fig.2(a) the
longitudinal resistance $R_{2,3}$ for the upper edge, and $R_{6,5}$
for the bottom edge. For the parameters used in Fig.1(A1) with an
indirect negative gap, the coexistence of twisting edge band and
bulk band does not directly show a topological property. Two Hall
(longitudinal) resistances are very small and almost equal, because
that the system is translationally invariant. For the parameters used
in Fig.1(A3), though the edge band is already somewhat distorted with
two edge currents having different speeds, the measurement can give
no new information other than the normal topological insulator. The
Hall (longitudinal) resistances measured at different place (edge)
are the same. Within the gap, the Hall resistances give a quantized
plateau ($h/e^2$) characterized by the topological number $\nu=1$,
and two longitudinal resistances are zero, because of the absence of
back scattering.

For the parameters used in Fig.1(A2), the twisting edge band case,
the results are very different and interesting. When the Fermi
energy $E_{F}$ is within the gap but out of the range of the
twisting of edge band, all measurements still show normal
topological property by giving the plateau. When the Fermi energy
goes within the twisting area, the situation is totally changed.
Let us first look at the longitudinal resistance. We still
have $R_{6,5}=0$, because there is only one edge state $b$ on
the bottom edge of the sample, no back scattering is allowed there,
the voltage drop is zero with $V_{6}=V_{5}$.
However, on the upper edge, $R_{2,3}$ is nonzero and it is
about $0.25 h/e^2$. This is because we have three
edge states on the upper edge, two of them move to the right and the
other one moves to the left. As one pair of them moves in the opposite
directions, not topologically protected, the back scattering
is allowed. Thus, the voltage may drop, $V_{2}\neq V_{3}$ to give a
nonzero resistance $R_{2,3}$ on the upper edge. Specifically, as
lead-2 is on the left side of the lead-3, we have
$V/2=V_{1}>V_{2}>V_{3}> V_{4}=-V/2$.
The two Hall resistances also change and they are no longer equal to
the value $h/\nu e^2$, although the Chern number of the system is still
$\nu=1$. In particular, as $V_{2}\neq V_{3}$ and
$V_{6}=V_{5}$, we can see that the left side Hall resistance
$R_{2,6}=(V_{2}-V_{6})/I_{1}$ is no longer same as the right side Hall
resistance $R_{3,5}=(V_{3}-V_{5})/I_{1}$: $|R_{2,6}|$ is decreased
to about $0.7 h/e^2$ within the twisting-edge-band region but
$|R_{3,5}|$ is larger than $ 0.9 h/e^2$. It should be emphasized again
that the present system is translationally invariant. However, from
the results above, the Hall resistance does break the translational
invariance. This novel phenomenon, the breaking of the translational
invariance of the Hall resistance in a translationally invariant
system, origins from the twisting edge band and the combination of
the topological protected and unprotected edge states. This property
is unique to the topological system with a twisting edge band and can
not be observed in either normal topological insulators or
non-topological systems. In addition, we also witnessed the oscillation
of $R_{3,5}$ and $R_{2,3}$ for the parameters used in Fig.1(A2). This
is because of the Fabry-Perot interference between the lead-2 and
lead-3. The number of oscillation are determined by the distance
between them.

\begin{figure}
[t]
\begin{center}
\includegraphics[width=8cm]{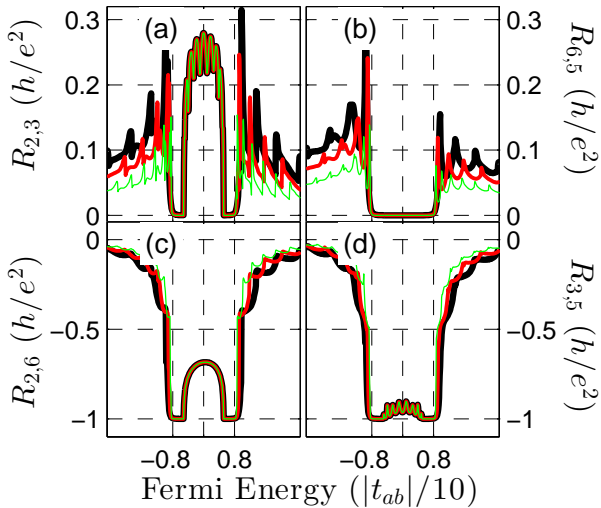}
\end{center}
\caption{\label{fig:Fig. 3} (Color online) For the parameters used
in Fig.1(A2), the resistances v.s. Fermi energy for sample widthes
$W=50a$ (the broadest black line), $60a$ (the red line), and $80a$
(the thinnest green line). }
\end{figure}

In order to confirm that the breaking of the translational invariance of
Hall resistance is due to the edge states, we show the Hall and
longitudinal resistances versus the width of the
sample in Fig.3. The change of width only has the effect on the
bulk bands and should not affect the edge bands when the sample is wide
enough. From Fig.3, it can be seen that outside the gap, all the
four resistances are changed when the width changes. However, within
the gap, the resistances maintain the same for different widthes.
It clearly shows that the breaking of the translational invariance of
the Hall resistance does come from the twisting edge band.

One may argue that the 6-lead measurement itself already breaks the translational invariance, as
the left Hall bar is close to the higher voltage side and the
right Hall bar is close to the lower voltage side \cite{addref3}.
Following we consider the 4-lead set-up of Hall resistance [see
the inset in Fig.4(A)] and vary the measurement position. In addition,
disorder effect is also studied. Let us
suppose the system having a uniform distributed Anderson disorder,
that does not break the translational invariance.
In the presence of disorder, the Hall resistance $|R_{2',4'}|$
increases with the measure position moving from the left to the
right [see Fig.4(A)]. This clearly implies that the Hall resistance
depends on the measure position, breaking the
translational invariance. In addition, on the left edge of the sample, the Hall
resistance is almost not affected by the disorder. When the sample
is long enough, the Hall resistance measured on the right edge is
close to $|R_{2',4'}|= h/\nu e^{2}$, quantized by the topological
number $\nu=1$.

\begin{figure}
[th]
\begin{center}
\includegraphics[width=8.35cm]{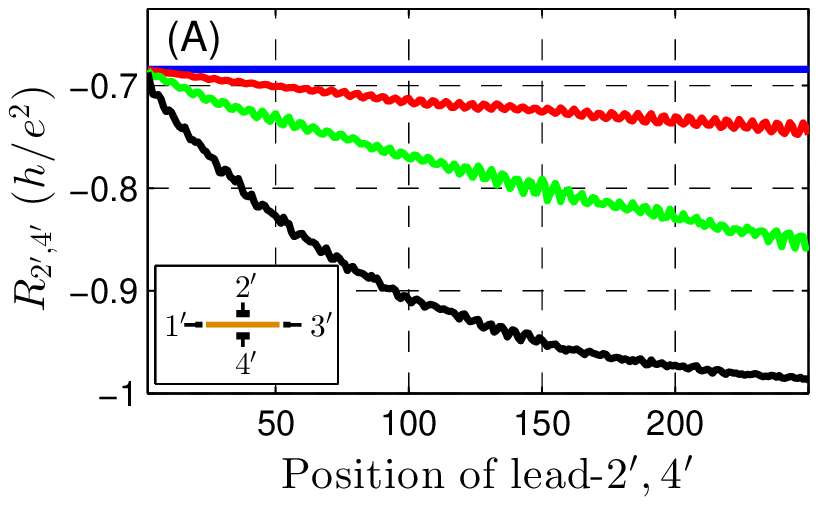}\\
\vspace{0.2cm}
\includegraphics[width=8.35cm]{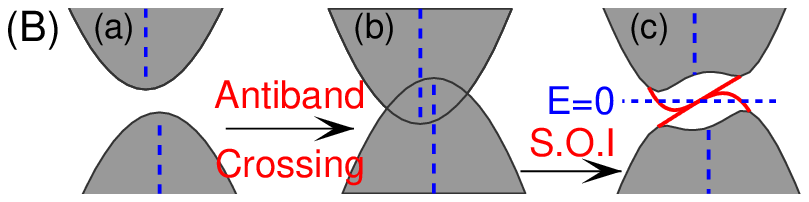}
\end{center}
\caption{\label{fig:Fig. 4} (Color online) (A). For the
parameters used in Fig.1(A2), the 4-lead measurement of Hall
resistances v.s. the position to measure at $E_{F}=0$. From top to bottom, the blue,
red, green, and black lines are for the disorder strength $Dis=0$,
$\Delta/8$, $\Delta/4$, and $\Delta/2$, respectively. Here the
gap is $\Delta=0.16|t_{ab}|$. The results are calculated
with the width of sample $W=70a$, by the average of 700 disorder
configurations. (B). The schematic diagram
of another method to realize twisting edge bands.}
\end{figure}

Finally, we should point out that though our results are obtained from an
ideal model, the twisting edge bands can be found in some real
systems. For example, supposing we initially have the two band
system as shown in Fig.4(B-a), whose symmetry axis of upper band is shifted
from that of the lower band. Then with the anti-band crossing
[Fig.4(B-b)], the pseudo spin-orbital interaction may open a gap and
leads to a twisting edge band [Fig.4(B-c)].

\emph{In conclusion,} we have shown that, with a twisting edge
bands, the system has both the topological protected and
unprotected edge states. In such a system, the Hall resistance is
not determined by the topological number alone. In particular, the Hall resistance
depends the measure position even for a translationally invariant
system.

\emph{Acknowledgments:} This work was financially supported by NBRP
of China (2012CB921303 and 2009CB929100), NSF-China under Grants Nos. 10821403,
10974236, and 11074174, and US-DOE under Grants No. DE-FG02-
04ER46124.


\begin{thebibliography}{99}
\bibitem{addref3}
Notice that this argument is not supported by the measurement for
both the topological trivial system and normal topological system.
\end{thebibliography}
\end{document}